\documentclass[reprint,aps,pra,amsmath]{revtex4-1}
\usepackage{newtxtext, newtxmath} 
\usepackage{graphicx, braket, siunitx}
\usepackage{dcolumn, multirow} 
\usepackage{hyperref, cleveref} 

\newcommand*\op[1]{\hat{#1}}
\newcommand*\U{\op{\mathcal{U}}}
\renewcommand*\H{\op{\mathcal H}}
\newcommand*\sx{\op\sigma_{\mkern-4mu x}}
\newcommand*\sy{\op\sigma_{\mkern-3mu y}}
\newcommand*\sz{\op\sigma_{\mkern-4mu z}}
\renewcommand*\sp{\op\sigma_{\mkern-4mu+}}

\renewcommand*\a{\op a}
\let\Im\undefined\DeclareMathOperator\Im{Im}

\begin{document}

\author{Jake Lishman}
\affiliation{Blackett Laboratory, Imperial College London, %
             London SW7 2AZ, United Kingdom}
\author{Florian Mintert}
\affiliation{Blackett Laboratory, Imperial College London, %
             London SW7 2AZ, United Kingdom}
\title{Trapped-Ion Entangling Gates Robust Against Qubit Frequency Errors}
\date\today

\begin{abstract}
Entangling operations are a necessary tool for large-scale quantum information processing, but experimental imperfections can prevent current schemes from reaching sufficient fidelities as the number of qubits is increased.
Here it is shown numerically how multi-toned generalizations of standard trapped-ion entangling gates can simultaneously be made robust against noise and mis-sets of the frequencies of the individual qubits.
This relaxes the degree of homogeneity required in the trapping field, making physically larger systems more practical.
\end{abstract}

\maketitle

\section{Introduction}

A major goal in quantum information processing is to reach the level of a fast, highly-scalable universal quantum computer.
A device at this level is proven to have computational capabilities for certain classes of problems which exceed any possible classical computer~\cite{Shor1994,Bernstein2017}, and would have major applications in a broad range of fields spanning all the physical and computational sciences~\cite{Lanyon2010a,Peruzzo2014a,Dunjko2016}, making an inherently quantum world accessible to simulation and investigation.
Several physical technologies are being developed in parallel in search of this target~\cite{Makhlin1999,Morley2013,Kok2007}, of which trapped ions are commonly recognized as one of the two leading platforms, along with superconducting qubits~\cite{Krantz2019,Bruzewicz2019}.
To reach universality for a constant number of qubits, only a small set of operations is absolutely required: a small number of single-qubit operations, and a single two-qubit entangling operation.

Throughout their development, quantum gate implementations have always contended with noise reduction, with varying estimates placing the maximum allowable probability of failure per gate at between \num{e-2} and \num{e-4}~\cite{Knill2005}.
Single-qubit gates have been achieved in ion traps at fidelities over \SI{99}{\percent} for over a decade~\cite{Benhelm2008}, with more recent works taking the average gate infidelity to \num{e-6}~\cite{Harty2014}.
The current state-of-the-art fidelities for two-qubit gates are performed in ion traps, achieving infidelities of less than \num{e-3}~\cite{Gaebler2016,Ballance2016} with laser-induced gates, and \num{3e-3} with microwave-controlled schemes~\cite{Harty2016}, requiring very low tolerances in homogeneity and stability of control and trapping fields, with scalability remaining a large problem.
Proposals to enlarge ion-trap computers typically focus on producing modular systems, and either on physically shuttling ions~\cite{Lekitsch2015} or introducing probabilistic photonic interconnects between separated traps~\cite{Monroe2014}.
Both of these methods exacerbate existing sources of noise by increasing either the physical distance or the number of external supplies and bulk optics that field coherence must be maintained across.
It is still imperative that entangling operations can be achieved that are robust against degraded conditions and controls.

These two primary qubit technologies in ion traps typically suffer from differing dominant degradation effects, though their methods of action are similar.
In both the optical and microwave regimes, the qubits are separated by too great distances due to their Coulomb repulsion to interact directly, but this same force can be used to engineer an interaction using the shared motion as a temporary bus mode~\cite{Sorensen1999}.
The reliance on the motion creates another potential source of infidelity, alongside the necessity of keeping the qubit frequencies entirely coherent with each other and the driving fields.
In ion-trap gates, these noise sources are typically macroscopic components; lab temperature and electrode voltage drifts decohere the motional mode, while long-term laser- and microwave-field frequency and amplitude fluctuations primarily affect qubit frequency splittings.

For ion-trap qubits, there has been significant interest in making gates resilient against unwanted heating and frequency errors of the bus mode using multi-toned driving fields~\cite{Webb2018,Shapira2018}, or by amplitude or phase modulation~\cite{Zarantonello2019,Milne2020}.
Early microwave-controlled gates necessitated dynamical-decoupling methods to protect against overall fluctuations in the magnetic field~\cite{Bermudez2012,Harty2016}, with more recent proposals for hyper-fine qubits considering gate speed-ups by inserting coupling to more motional modes~\cite{Arrazola2018}, or experimental simplifications by decoupling from global qubit frequency mis-sets and oscillations without additional fields~\cite{Sutherland2019}.

The scheme illustrated here extends the previous literature by using a multi-tone extension to the M\o lmer--S\o rensen scheme to produce a gate resilient against all frequency errors on one or both of the qubits individually.
This scheme is applicable to all ion-trap qubit encodings, including magnetic-field-sensitive optical qubits, and produces an improvement in infidelity around the current threshold of error-correction of over two orders of magnitude, without being specifically generated for any particular offset magnitude.
The same numerical optimization methods can be applied to produce a driving scheme that minimizes the average infidelity for any error model, as the errors are considered non-perturbatively.
It may also be implemented simply in experiments, requiring no fields to be added; an arbitrary waveform generator is the sole necessity over the original M\o lmer--S\o rensen implementation~\cite{Sackett2000}.

\section{Model}

\subsection{System}

\begin{figure}%
    \includegraphics{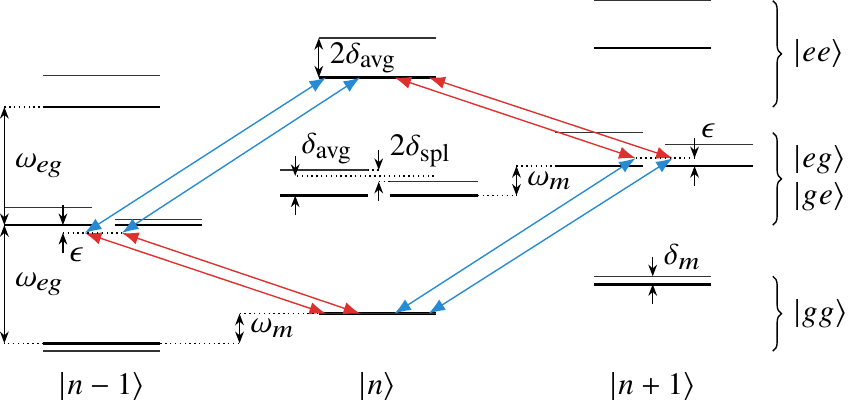}%
    \caption{\label{fig:errorlevels}%
        The energy levels of the standard M\o lmer--S\o rensen gate operation in the presence of frequency errors.
        Thick black lines denote the expected energy levels, whereas thin ones show the modified structure, and the driving is marked in red and blue for the appropriate sideband.
        An error $\delta_m$ in the motional frequency $\omega_m$ causes phonon levels to shift, but maintains resonance of all transition paths.
        Any error in qubit frequencies causes the two-photon process to be off-resonant for some starting states; a shift in the average of the carrier frequencies $\delta_{\text{avg}}$ changes the energy gap of $\ket{gg}\leftrightarrow\ket{ee}$ by $2\delta_{\text{avg}}$ while leaving $\ket{ge}\leftrightarrow\ket{eg}$, and an energy splitting between the two qubits $2\delta_{\text{spl}}$ has the opposite effect.
}%
\end{figure}

The system Hamiltonian for two harmonically-trapped ions considering only a single motional mode is
\begin{equation}\label{eq:hamiltonian-system}
    \H_{\textsc s}/\hbar = \frac12\Bigl(\bar\omega_{eg}^{(1)}\sz^{(1)} + \bar\omega_{eg}^{(2)}\sz^{(2)}\Bigr) + \bar\omega_m \a^\dagger\a,
\end{equation}
where $\bar\omega_{eg}^{(n)}$ is the qubit frequency separation of the $n$th ion and $\bar\omega_m$ is the frequency of a phonon of motion that has $\a$ and $\a^\dagger$ as annihilation and creation operators.
For ideal gate operation, the two separate qubit frequencies should be equal and all frequencies should be exactly known.
In reality, however, several noise sources conspire to modify these values over the course of a complete experiment, and the true frequency $\bar\omega$ is formed of a known component $\omega$ with the addition of some deviation $\delta$ as $\bar\omega = \omega + \delta$.
Modifications to the motional frequency $\delta_m$ occur primarily due to endcap voltage drifts, causing apparent dephasing effects when averaged over several gate realizations.
The dominant sources of error on the $n$th qubit frequency $\delta_{eg}^{(n)}$ depend strongly on the encoding of the qubits; magnetic-field-sensitive qubits will generally suffer most from local variations in the field, while the frequency separation of optical qubits is more commonly mis-set due to slow drift of the spectroscopy laser.
The errors on the two qubit frequencies can be parametrized as individual differences from the estimated frequency, as in \cref{eq:hamiltonian-system}, but it is more convenient for the analysis to consider an error in the estimation of the average frequency $\delta_{\text{avg}} = (\delta_{eg}^{(1)} + \delta_{eg}^{(2)})/2$ and the distance of each individual value from this average $\delta_{\text{spl}} = (\delta_{eg}^{(1)} - \delta_{eg}^{(2)})/2$.
These frequencies are diagrammed in the context of the energy-level scheme in \cref{fig:errorlevels} for the standard M\o lmer--S\o rensen gate detuned from the sidebands by an amount $\epsilon$.

Moving to a rotating frame defined by $\U = \exp(i\H_{\textsc s} t/\hbar)$, the driving--ion interaction Hamiltonian is
\begin{equation}
\label{eq:hamiltonian-interaction-picture}
\begin{aligned}
    \H_{\textsc i}/\hbar = {}
        &\tilde f(t) e^{i(\omega_{eg} + \delta_{\text{avg}})t}
        \Bigl(e^{i\delta_{\text{spl}} t}\sp^{(1)}
              + e^{-i\delta_{\text{spl}} t}\sp^{(2)}\Bigr)\\
        &\qquad\times\Bigl(1 + i\eta e^{i(\omega_m + \delta_m)t} \a^\dagger
                        + i\eta e^{-i(\omega_m + \delta_m)t} \a\Bigr)\\
        &\!{}+ \text{H.c.},
\end{aligned}
\end{equation}
where the Lamb--Dicke condition that $\eta\sqrt{2n+1}\ll1$ has been assumed, and terms oscillating at the same order as $\omega_{eg}$ are omitted.
The complete driving term $\tilde f(t)$ is ${\tilde f(t) = \sum_j f_j(t) \exp\bigl(-i(\omega^{(j)}_s + \delta^{(j)}_s)t\bigr)}$ and comprises of two terms: the latter sideband-selection term and a slowly oscillating driving term $f_j$, such that each term in \cref{eq:hamiltonian-interaction-picture} is of a comparable frequency to the acoustic trap frequency.
The selection frequency $\omega_s$ is set to $\omega_{eg} + n\omega_m$ to pick out the $n$th sideband, where $n=0$ is the carrier, $n=1$ is the blue sideband, and $n=-1$ is the red sideband.
The sideband-driving term $f_j(t)$ has frequency components that are small compared to the sideband separation frequency $\omega_m$, so that only the targeted sideband is excited.
For the M\o lmer--S\o rensen gate, the ions are globally illuminated by a blue field with a selection frequency $\omega_s^{(b)} = \omega_{eg} + \omega_m$ using a constant-amplitude slightly off-resonant drive $f_b(t) = \Omega e^{i\epsilon t}$, simultaneously with a red field at $\omega_s^{(r)} = \omega_{eg} - \omega_m$ and $f_r(t) = f_b^*(t)$, leading to a Hamiltonian
\begin{equation}
\label{eq:hamiltonian-molmer-sorensen}
\H_{\text{\sc ms}} = -\eta f_b(t) e^{i\delta_m t}\a^\dagger\cdot\mathopen{}\left(
    \begin{alignedat}2
     &\cos&\bigl((\delta_{\text{avg}}+\delta_{\text{spl}})t\bigr)&\sy^{(1)}\\
    +&\sin&\bigl((\delta_{\text{avg}}+\delta_{\text{spl}})t\bigr)&\sx^{(1)}\\
    +&\cos&\bigl((\delta_{\text{avg}}-\delta_{\text{spl}})t\bigr)&\sy^{(2)}\\
    +&\sin&\bigl((\delta_{\text{avg}}-\delta_{\text{spl}})t\bigr)&\sx^{(2)}\\
    \end{alignedat}\right) + \text{H.c.}.
\end{equation}
In the absence of errors, this degrades to the standard Hamiltonian $\H_{\text{\sc ms}} = -\eta\bigl(f_b(t)\a^\dagger + f_b^*(t)\a\bigr)\bigl(\sy^{(1)} + \sy^{(2)}\bigr)$.
In \cref{eq:hamiltonian-molmer-sorensen}, the two selection error terms $\delta_s^{(r)}$ and $\delta_s^{(b)}$ have been reparametrised to average and splitting terms with the same treatment as the qubit error terms.
In this form, the splitting term appears only as an addition to the motional detuning $\delta_m$, while the average similarly modifies the average qubit detuning $\delta_{\text{avg}}$, allowing these two pre-existing terms to completely represent static mis-sets in the selection frequencies.

\Cref{eq:hamiltonian-molmer-sorensen} is analytically solvable only when no qubit frequency errors are present, resulting in a time-evolution operator
\begin{equation}\label{eq:propagator-molmer-sorensen}\begin{aligned}
\U_{\text{\sc ms}}(t) = {}
    &\op{\mathcal D}\Bigl(
        \bigl(\sy^{(1)}+\sy^{(2)}\bigr){\textstyle \int_0^t} f_b(t_1)\,\mathrm dt_1
    \Bigr)\\
    &\quad\times\exp\Bigl(
        2i\sy^{(1)}\sy^{(2)}\Im {\textstyle \int_0^t\mathrm dt_1\int_0^{t_1}\mathrm dt_2}\, f_b(t_1)f_b^*(t_2)
    \Bigr),
\end{aligned}\end{equation}
where $\op{\mathcal D}$ is the motional phase-space displacement operator $\op{\mathcal D}(\alpha) = \exp(\alpha\a^\dagger - \alpha^*\a)$.
As such, the first term defines the coupling of the qubits individually to the excitation of the motional mode, and the second term represents a true entangling interaction between the two qubits.
Together, these two terms form two conditions that must be satisfied simultaneously at the gate time: the motional phase-space displacement must return to zero; the qubit entanglement phase accumulation must reach the desired level.

The driving function $f_b(t)$ cannot be designed to eliminate error terms from the complete Hamiltonian, and with their effects active, an exact time-evolution operator like \cref{eq:propagator-molmer-sorensen} cannot be found.
Series-expansion methods neither truncate nor converge in a computable number of steps; the non-commutation of the Pauli operators $\sigma_x$ and $\sigma_y$, along with dependence on increasingly large motional excitations at higher orders frustrate the Magnus and similar expansions, while the aperiodicity of the system prevents a reasonable Floquet approach.
Instead, numerical techniques are used here to access and minimize the gate's response to errors.

\subsection{Optimization}

It is first important to quantify how performant a quantum gate is, so that some form of improvement can be found and optimized.
Any meaningful measure of the success of an operation must take into account all possible states that the system may exist in.
One such measure is the gate infidelity, defined by $I = 1 - \sum_k {\bigl\lvert\braket{\psi_k|\U_{\text{\sc ms}}(\vec\delta)|\chi_k}\bigr\rvert}^2 / K$, in terms of all types of detunings $\vec\delta$ and $K$ pairs of start and ideal target states $\{\ket{\chi_k}\!,\,\ket{\psi_k}\}$ respectively where the start states span the Hilbert space concerned.
To ensure that the gate is resilient to detunings with a wide range of magnitudes, an appropriate figure of merit is an expectation of the total infidelity $E[I(\vec\delta)] = \int\!I(\vec\delta)\,\mathrm{d} w(\vec\delta)$, for a suitable weight function $w$; typically this can be taken as an adequately-dimensioned normal distribution as a reasonable proxy for experimental uncertainties.
This modified target causes the optimizer to prefer parameters which provide good fidelities over a range of errors, with a hyper-parameter $\vec\sigma_\delta$, being the standard deviations of the error distributions, affecting how heavily larger errors are weighted.
Notably, the use of an expectation does not require the optimal schemes to have perfect infidelity at zero error, but a suitable choice of weight function may ensure that any remnant error will be negligible.

The optimizations presented here will consider a shaped driving field $f(t)$---the subscript $b$ is dropped for simplicity---with multiple frequency components (``tones'') simultaneously to minimize the effects of the error terms on the final gate operation, taking $f(t) = \sum_{k=1}^{n} c_{n,k} e^{ik\epsilon t}$ where the $c_{n,k}$ are complex variables with dimensions of frequency.
The targeted figure of merit is the expectation of gate infidelity, averaged over all possible electronic starting states weighted equally and over all possible detunings weighted as a normally distributed error model.
An understanding of the precise details of the numerical methodology is not necessary to appreciate the subsequent results, so further discussion is deferred to \cref{sec:optimisation-methods}.

\section{Results}

\begin{table*}
    \newcommand*\spacingstrut{\rule{0pt}{3ex}}
    \newcolumntype{x}{D..{1.3}}
    \begin{ruledtabular}\begin{tabular}{crrrrxxxxxx}
    Tones & $\tau_n/\tau_1$ & $\delta\!f_n$ & $\epsilon_n$ & &
        \multicolumn1r{$c_{n,1}$} & \multicolumn1r{$c_{n,2}$} & \multicolumn1r{$c_{n,3}$} & \multicolumn1r{$c_{n,4}$} & \multicolumn1r{$c_{n,5}$} & \multicolumn1r{$c_{n,6}$} \\\hline
    \multirow2*2 & \multirow2*{3.368} & \multirow2*{0.132} & \multirow2*{1.188} & \spacingstrut
        $\lvert c\rvert$ &  0.066 &  0.934 &&&& \\
    &&&&$\phi/\pi$       & -0.032 &  0     &&&& \\
    \multirow2*3 & \multirow2*{3.185} & \multirow2*{0.033} & \multirow2*{1.256} & \spacingstrut
        $\lvert c\rvert$ &  0.103 &  0.979 &  0.090 &&& \\
    &&&&$\phi/\pi$       & -0.005 & -0.003 &  0     &&& \\
    \multirow2*4 & \multirow2*{4.836} & \multirow2*{0.555} & \multirow2*{0.827} & \spacingstrut
        $\lvert c\rvert$ &  0.051 &  0.405 &  0.539 &  0.359 && \\
    &&&&$\phi/\pi$       & -0.609 & -0.817 &  0.108 &  0     && \\
    \multirow2*5 & \multirow2*{4.542} & \multirow2*{0.622} & \multirow2*{0.881} & \spacingstrut
        $\lvert c\rvert$ &  0.048 &  0.450 &  0.516 &  0.414 &  0.183 & \\
    &&&&$\phi/\pi$       & -0.899 & -0.930 &  0.045 & -0.242 &  0     & \\
    \multirow2*6 & \multirow2*{6.529} & \multirow2*{0.482} & \multirow2*{0.613} & \spacingstrut
        $\lvert c\rvert$ &  0.055 &  0.098 &  0.413 &  0.733 &  0.215 &  0.128 \\
    &&&&$\phi/\pi$       & -0.616 & -0.785 & -0.954 &  0.007 & -0.043 &  0     \\
    \end{tabular}\end{ruledtabular}
    \caption{\label{tab:qubit-error-schemes}
        Tabulated values of coefficients for the multi-tone driving of the M\o lmer--S\o rensen gate optimized to reduce the effects of static qubit frequency errors.
        The driving field with $n$ tones takes the form $f_n(t) = \sum_{k=1}^n \lvert c_{n,k}\rvert e^{i\phi_{n,k}} e^{ik\epsilon_n t}$, where the units are scaled such that $\epsilon_1 = 4$ and $c_{1,1} = 1$ in the base case, and all driving fields have the same peak power usage.
        The gate time $\tau_n$ is given in terms of the standard gate, which has constant power, while the multi-tone gates have a maximum variation in the power of $\delta\!f_n$.
        The last phase is chosen as zero for all pulses; driving fields are equivalent up to a global phase.
    }
\end{table*}

\begin{figure}%
    \includegraphics{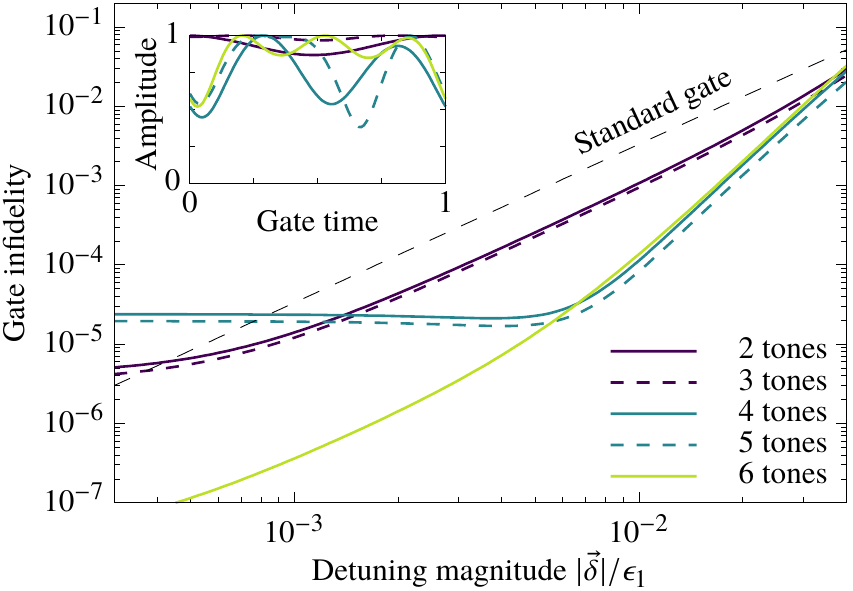}%
    \caption{\label{fig:multi-tone}%
        Main figure: gate fidelity for optimized driving schemes compared to the standard M\o lmer--S\o rensen scheme.
        Inset: total drive amplitude during the gate operation.
        The detuning error considered is in the ratio $\delta_{\text{avg}} = 2\delta_{\text{spl}}$. An error which causes the single-tone gate to leave the error-correction threshold of \SI{99.9}{\percent} causes an infidelity of only \num{2.5e-5} when four or more tones are used.
        The two- and three-tone gates are minor modifications of the standard driving, yet produce a three- to four-times improvement over the range of meaningful infidelities.}%
\end{figure}

We perform optimizations using the multi-tone parametrization of the driving field to produce gates resilient against all forms of static errors on the qubit frequencies and an average offset in the sideband-selection frequencies.
The maximum peak power usage of the interrogation source is fixed across all numbers of tones so as to form a fair comparison with the base gate, while the gate time is allowed to vary to facilitate this by making the base detuning $\epsilon$ a control parameter.
Aside from this detuning of the closest tone to the sideband, the other optimization variables are the relative strengths and phases of the tones in the driving field.

In \cref{fig:multi-tone}, the best driving schemes obtained are compared to the performance of the base gate at varying qubit detunings.
Due to the nature of any numerical optimization, and as the optimization landscape is infinite and non-periodic, it is impossible to ascertain if a true global maximum has been found.
However, sampling the initial parameter space increasingly finely can arbitrarily reduce the possibility of having missed a better result.
The results presented here then are most correctly lower bounds on the maximum achievable fidelities.
The hyper-parameter $\vec\sigma_\delta$ was chosen to prioritize the minimization of infidelity
for detunings of such a magnitude that the base gate is close to, but not quite in, the error-correcting region.
This prioritizes cases where qubit frequency errors would prevent current gates from being computationally viable, and is largely unconcerned with situations where such errors would not be the dominant terms.
At lower detuning magnitudes, the monotone gate is able to out-perform these numerical schemes, but only in regions where the error is insignificant.

The driving fields resulting from these optimizations are specified in \cref{tab:qubit-error-schemes}, and their time-dependent amplitudes are shown in the inset of \cref{fig:multi-tone}.
For two- or three-tone gates, the optimized drivings are minor perturbations of a standard gate performed with two loops in phase-space, with maximum relative amplitude variation of $0.132$ and $0.033$ respectively, but the largest improvements are seen once four tones are included.
For errors which cause the base gate to have fidelities on the thresholds of the error-correcting regions, \SI{99}{\percent} and \SI{99.9}{\percent} depending on the particular definition, a four-tone gate using the same amount of peak power has infidelities of \num{1.0e-3} and \num{2.5e-5} respectively---$10$ and $250$ times smaller.
This does, however, come at a cost in gate time; this gate requires slightly under five times the amount of time to complete, largely because of the increase of the number of loops completed in phase space.

The infidelity of the base gate varies predominantly quadratically with a change in the magnitude of the qubit errors.
We have found numerically that a minimum of four tones are required to improve this scaling behavior over any sizable region of interest; \cref{fig:multi-tone} shows this improvement in the scaling through steeper gradients for four and higher numbers of tones.
This new scaling is quartic for realistic errors, although a new constant offset is introduced at lower magnitudes.
The minimum infidelity is then non-zero for the optimized gates, however it can be made negligibly small with the addition of further tones.
As this method also allows the easy selection of the region of interest, this does not pose any hard limit of fidelity from this multi-tone driving.

\begin{figure}%
    \includegraphics{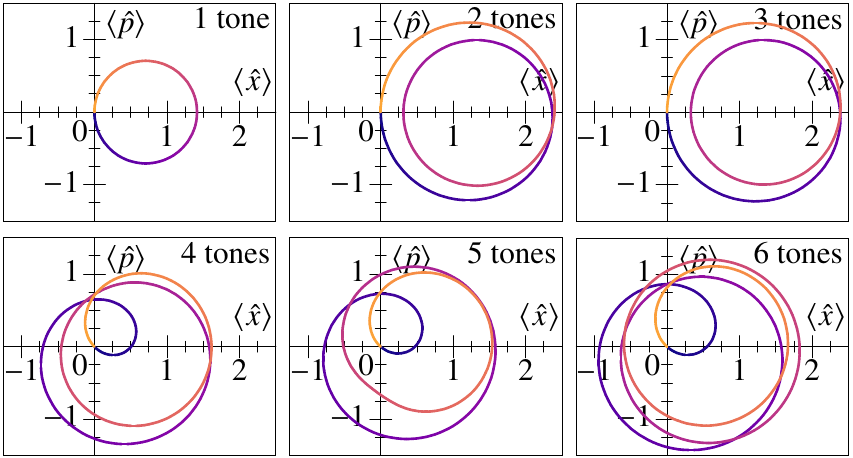}%
    \caption{\label{fig:phase-space}%
        Motional phase-space trajectories of the different multi-tone gates also plotted in \cref{fig:multi-tone},  with the same peak power usage and different gate times.
        Structural changes to the trajectories only occur on even numbers of tones.
        The relative time through the gate is represented by the line color, moving from purple (dark) to orange (light).
        Valid qubit-phase advancements are $(4n+1)\pi/4$ for integer $n$; the single-tone gate has $n=0$, while two to five tones have $n=1$ and six tones has $n=2$.
    }%
\end{figure}

It is notable that only even numbers of tones make significant changes to the fidelity response of the gate; \cref{fig:phase-space} shows that structural changes to the phase-space trajectories only occur at these points, despite the amplitude modulation being rather different between members of each even--odd pair.

\section{Optimization Methods}
\label{sec:optimisation-methods}

\subsection{Evaluation of Figure of Merit}

A successful optimization over several parameters typically requires hundreds of evaluations of the figure of merit, and the number of repetitions needed to adequately sample the initial parameter possibilities can easily push this to millions.
As the number of free parameters increases, so too does the average number of evaluations needed for convergence.
This can easily place restrictions on the driving fields that can be considered, or severely limit the exploration of the optimization landscape if the calculation complexity is not carefully attended to.

The inclusion of an integral in the figure of merit poses a particular speed concern; numerical integration must always evaluate the integrand several times, and each evaluation requires a complete numerical solution to the Schr\"odinger equation.
The number of operations required to achieve a certain precision generally scales exponentially with the dimensionality of the integral, mandating that the integrand should be sampled in minimal locations.
In one dimension, integrals over a weight function can be evaluated to a high degree of accuracy with few abscissae using Gaussian quadrature.
The integrand is considered in terms of a polynomial set orthogonal under the real inner product $\langle f,\,g\rangle = \int\!f(x)g(x)\,\mathrm{d}w(x)$, allowing the integration of $I$ accurate to degree $2n-1$ to be expressed as a sum $\sum_{i=1}^n w_i I(x_i)$, where the $x_i$ are the roots of the $n$th-order polynomial and the weights $w_i$ are precalculated~\cite{Press2007}.
For a weight function of the form $w(x) = \exp\bigl(-x^2\bigr)$, as is the case for normally-distributed errors, the relevant polynomial set is the Hermite polynomials.
This method does not require recursive subdividing of integration regions to reach a desired accuracy unlike simpler Newton--Cotes schemes, reducing the total number of evaluations, and can also handle infinite regions without truncation.
Similar methods allow the extension to $d$ dimensions with better performance than the na\"ive $n^d$ achieved by nesting, although the lack of well-defined orthogonal polynomials does not permit generic constructions to arbitrary degree~\cite{Stroud1971}.

The evaluation of the figure of merit is also accelerated by considering symmetry of the integrand under transformations of the detunings and spanning basis.
After the absorption of the selection errors into other terms, the three remaining errors specified in \cref{eq:hamiltonian-molmer-sorensen} manifest differently throughout the action.
Shifts in the motional frequency leave all two-photon processes on-resonance, but modify the true gate time.
The qubit errors cause certain transitions to become energetically mismatched; a shift in the average carrier transition causes the $\ket{gg}\leftrightarrow\ket{ee}$ flopping to have an energy difference of $2\delta_{\text{avg}}$ from the sum of the red and blue photons required, but without lifting of the energy degeneracy of the $\ket{eg}$ and $\ket{ge}$ levels, the blue--blue and red--red processes which mediate entanglement between this manifold remain favorable albeit with a modified detuning from the virtual levels.
If instead the average is well-known but there are different carrier frequencies, the $\ket{eg}\leftrightarrow\ket{ge}$ transition cannot be on-resonance, but the blue--red process to promote $\ket{gg}$ to $\ket{ee}$ is.

This similarity can be quantified by considering how the evolution of the system changes when its initial state is modified by a time-independent unitary operator $\op{\mathcal V}$, such as $\sy^{(1)}$ which maps $\ket{gg}$ to $-i\ket{ge}$.
When a Hamiltonian $\H$ satisfying the Schr\"odinger equation $i\partial_t\U = \H\U$ is modified by $\op{\mathcal V}$ to $\H' = \op{\mathcal V}^\dagger\H\op{\mathcal V}$, the resultant time-evolution operator is $\U'=\op{\mathcal V}^\dagger\U\op{\mathcal V}$.
With the qubit error terms in \cref{eq:hamiltonian-molmer-sorensen} as explicit arguments, taking $\op{\mathcal V} = \sy^{(1)}$ leads to $\H_{\text{\sc ms}}'(\delta_{\text{avg}}, \delta_{\text{spl}}) = \H_{\text{\sc ms}}(-\delta_{\text{spl}}, -\delta_{\text{avg}})$, while $\op{\mathcal V} = \sy^{(2)}$ makes $\H_{\text{\sc ms}}'(\delta_{\text{avg}}, \delta_{\text{spl}}) = \H_{\text{\sc ms}}(\delta_{\text{spl}}, \delta_{\text{avg}})$.  We therefore find that $\delta_{\text{avg}}$ and $\delta_{\text{spl}}$ have equivalent effects on different starting states, and cause equal infidelities when totaled over the complete basis of gate operation.
Any shaped driving function $f(t)$ which minimizes a total gate error for an offset in the average qubit frequency will consequently also minimize the error due to a splitting between the two.
Further, the oscillations $\ket{gg}\leftrightarrow\ket{ee}$ and $\ket{eg}\leftrightarrow\ket{ge}$ are symmetric with respect to exchange of starting state if the sign of both errors simultaneously flip, \textit{i.e.} the dynamics of the transition $\ket{gg}\rightarrow(\ket{gg}-i\ket{ee})/\sqrt2$ exhibits the same infidelity behavior for $\delta_{\text{avg}}$ and $\delta_{\text{spl}}$ as $\ket{ee}$ does for $-\delta_{\text{avg}}$ and $-\delta_{\text{spl}}$.
In tandem, these two symmetries allow complete information of the average fidelity to be obtained by considering only half the possible initial states, thus taking half the time.

\subsection{Power-Usage Constraints}

Unlike the standard M\o lmer--S\o rensen scheme, the $n$-tone driving $f_n(t) = \sum_{k=1}^nc_{n,k}e^{ik\epsilon_n t}$ considered here has variable power usage ${\propto}{\lvert f_n \rvert}^2$ throughout the gate.
The supremum location for an arbitrary number of tones with given control parameters is calculated by reformulating the natural maximization problem into one of polynomial root-finding, which can be solved by eigenvalue methods on a companion matrix~\cite{Press2007}.
All extrema of the power constraint are located at the zeroes of the derivative $\partial_t{\lvert f_n(t)\rvert}^2$, which can be recast via multiplication by the non-zero term $\exp\bigl(i(n-1)\epsilon_nt\bigr)$ into the complex polynomial in $z=e^{i\epsilon_nt}$
\begin{equation}
\sum_{\substack{k=0 \\ k\ne n-1}}^{2n-2}
    \biggl(\sum_j c_{\vphantom kj}^{\vphantom*} c^*_{j-k+n-1}\biggr)
    \bigl(k-n+1\bigr)
    z^k
    = 0,
\end{equation}
where $j$ runs from $1$ to $k+1$ for $k < n-1$ and from $k-n+2$ to $n$ for $k > n -1$.
The roots $z_\ell$ are related to the temporal locations of extrema $t_{\ell,m}$ by $\epsilon_nt_{\ell,m} = \arg(z_\ell) - i\ln\lvert z_\ell\rvert + 2\pi m$, where the integer $m$ denotes the period of the driving, and the only roots of interest are in the first period and real, where $\lvert z_\ell\rvert = 1$ and $m = 0$.
The peak power usage follows simply by testing the $2n-2$ or fewer abscissae to find the global maximum.

The optimizations presented in the paper are performed using a standard \emph{unconstrained} \textsc{bfgs} algorithm~\cite{Press2007} over the free ratios $\lvert c_{n,k}/c_{n,1} \rvert$, the relative phases $\phi_{n,k}$ and the principle detuning $\epsilon_n$, which can vary entirely freely.
The constraint that the peak power usage is equal under both schemes is then achieved by fixing the absolute value of $c_{n,1}$ such that $\max_t{\lvert f_n(t)\rvert}^2 = \max_t{\lvert f_1(t)\rvert}^2$, inside the figure of merit calculation.
The final free parameter---the coupling strength of the base gate---is chosen to be $c_{1,1} = \epsilon_1 / 4$ to coincide with the shortest possible single-tone gate.

\subsection{Agreement with Prior Results}

In order to gauge the reliability of the numerical method, a comparison can be made with analytically constructed control solutions, such as the shaped pulses rendering gates robust against errors solely in the motional frequency.
The optimal pulse shapes found with this method reproduce those previously reported~\cite{Webb2018,Shapira2018}, which are significantly different to those presented in \cref{tab:qubit-error-schemes} and illustrated in \cref{fig:multi-tone,fig:phase-space}.
In particular, the average absolute phase-space displacement is kept as close as possible to zero~\cite{Haddadfarshi2016} to lessen the effects of thermal fluctuations and trap-frequency offsets, whereas this is not the case for qubit-frequency errors.

These methods have reproduced the analytically known optimum solutions in situations where they are known to exist, but have also been numerically shown to produce highly robust gates in qualitatively different situations where a pen-and-paper construction is not possible, highlighting the versatility of the methodology.

\section{Conclusions}

Simply-synthesized multi-tone drivings can massively reduce errors due to qubit frequency shifts on one or both qubits simultaneously in the standard M\o lmer--S\o rensen gate, without increasing the amount of peak power required.
With four or more tones, the quadratic scaling of the infidelity with respect to the qubit error size can be improved to fourth-order, with a constant maximum fidelity ceiling which is raised by the addition of further tones.
This method is not unique to any method of driving nor qubit encoding, and can be applied universally across all standard trapped-ion processors with little-to-no additional hardware required, but is most useful in the field of optical qubit systems where previous microwave techniques do not readily apply.
The techniques used to quickly numerically optimize pulse sequences with a minimum number of simulations and to apply the strictly non-linear power usage constraints are general, applicable to all numerical infidelity optimizations.

\begin{acknowledgments}
We are grateful for stimulating discussions with Oliver Corfield, Jacopo Mosca Toba, Mahdi Sameti, Fr\'ed\'eric Sauvage, Richard Thompson, and Simon Webster.
Financial support by EPSRC through the \textit{Training and Skills Hub in Quantum Systems Engineering} EP/P510257/1 is gratefully acknowledged.
\end{acknowledgments}

\end{document}